\pdfoutput=1
\documentclass[11pt]{article}
\usepackage[a4paper,margin=2.5cm]{geometry}
\usepackage{mathptmx}
\usepackage[expansion=false]{microtype}
\usepackage{amsmath,amssymb}
\usepackage{graphicx}
\usepackage{booktabs}
\usepackage[numbers,sort&compress]{natbib}
\usepackage{enumitem}
\usepackage{xcolor}
\usepackage[hidelinks]{hyperref}
\usepackage[capitalise,noabbrev]{cleveref}
\usepackage{setspace}
\newtheorem{finding}{Finding}

\title{Post-Quantum Authentication of IoT Data on Ethereum:\\
Signatures, the Calldata Floor, and Device-Signed Merkle Roots}
\author{Ming-Fong Sie\\[1ex]
\small Master Program in Intelligent Computing and Big Data, Chung Yuan Christian University, Taoyuan, Taiwan\\
\small E-mail: \href{mailto:mingfong@cycu.edu.tw}{mingfong@cycu.edu.tw}}
\date{}

\begin{document}
\maketitle

\begin{abstract}
Field devices deployed today will still be producing records when quantum computers may break their signatures, so systems that anchor IoT data on blockchains need a post-quantum (PQ) authentication path now. We measure its cost on Ethereum under the current execution rules, including the EIP-7623 calldata floor and the EIP-7825 transaction cap, and under the calldata repricing scheduled for the Glamsterdam upgrade. We combine device-side measurements of ML-DSA, Falcon, SLH-DSA, LMS and XMSS, reproduced costs of open-source Solidity verifiers, a new standard-compliant LMS/HSS verifier for SHA-256 and SHA-256/192, and complete transactions on an Osaka-rule EVM, priced with today's verifiers and with proposed precompiles. Three results stand out. Signing every record costs 8.7 times as much as ECDSA even with an ML-DSA precompile, because the calldata floor rather than verification sets the price. When each device signs a Merkle root over 100 records, a standard LMS verifier available today costs about 3,350 gas per record, against 245 gas for ECDSA in the same design, and it is unaffected by the scheduled repricing, which raises floor-bound designs by up to 1.9 times. With the proposed precompiles, ML-DSA-44 costs 1,031 gas per record, and a hybrid ECDSA and ML-DSA signature adds only 2.5\%. Post-quantum device authentication is therefore deployable on Ethereum without protocol changes, provided that the latency of batching and the data-availability properties of the chosen variant fit the application.
\end{abstract}


\section{Introduction}
\label{sec:intro}

Blockchains are used to anchor data produced by IoT devices so that it can later serve as tamper-evident evidence, for instance of agricultural production backing a loan in supply chain finance~\cite{sie2026cost,wu2021mlav,sie2022secured}. The authenticity of each record rests on the signature of the device that produced it, today almost always an elliptic-curve signature. Shor's algorithm~\cite{shor1997polynomial} would let a cryptographically relevant quantum computer recover a device's signing key from its public key and forge new records in its name.

Records anchored before such a computer exists remain protected by their on-chain timestamps, provided that the chain's own history remains trustworthy, which depends on Ethereum's consensus signatures being migrated in time. The pressing problem is therefore the \emph{devices}: field units deployed now will still be producing records when quantum attacks become feasible, and their firmware and key infrastructure are hard to update remotely~\cite{mosca2018cybersecurity}. Systems deployed today should therefore be crypto-agile and able to authenticate records with post-quantum (PQ) signatures.

NIST has standardized ML-DSA~\cite{fips204} and SLH-DSA~\cite{fips205}, approved the stateful hash-based schemes LMS and XMSS~\cite{sp800208,rfc8554,rfc8391}, and is standardizing Falcon~\cite{falcon2020}. Ethereum verifies none of them natively; precompiles for ML-DSA~\cite{eip8051,eip8355} and Falcon~\cite{eip8052} are draft proposals not scheduled for any upgrade, and Solidity verifiers cost up to millions of gas~\cite{zknox_ethdilithium,zknox_ethfalcon,kysil2025poqeth}. PQ signatures are also one to two orders of magnitude larger than ECDSA signatures, and Ethereum now charges data-heavy transactions a calldata floor (EIP-7623~\cite{eip7623}), caps the gas of a transaction (EIP-7825~\cite{eip7825}), and will raise the floor further in the scheduled Glamsterdam upgrade (EIP-7976~\cite{eip7976,eip7773}).

Signing a Merkle root over a batch of messages is a classical way to amortize expensive signatures~\cite{merkle1987digital}, recently revisited for PQ signatures~\cite{aguilarmelchor2024batch}. We do not propose the construction. We measure what PQ authentication of IoT records costs on Ethereum, with and without such batching, and analyse the choices that determine the cost: what is posted, where keys are kept, which scheme is used, and which data-availability, liveness and latency properties each variant has.

\paragraph{Contributions.}
\begin{enumerate}[leftmargin=*]
\item \textbf{Device-side costs} of ML-DSA, Falcon, SLH-DSA, LMS, HSS and XMSS from C benchmarks on x86-64 and ARM64 cores and published Cortex-M4 measurements, with uplink bytes and memory per design (\cref{sec:res-device}).
\item \textbf{Verification costs}: reproduced costs of the ZKNOX ML-DSA and Falcon verifiers, and a new RFC 8554/SP 800-208-compliant LMS/HSS verifier for $n=32$ and $n=24$, executed for real, with its worst case derived from a regression on Winternitz chain steps (\cref{sec:res-solidity}).
\item \textbf{End-to-end costs under current and scheduled rules} for per-record signatures and device-signed roots, with replay protection, compared with ECDSA and P-256 in the same design, including hybrid ECDSA and PQ signatures (\cref{sec:res-pr,sec:res-dr,sec:res-keys,sec:res-sens}).
\item \textbf{A validated cost model} with closed-form break-even points, and an analysis of data-availability tiers and of alternatives to contract-level verification (\cref{sec:designs,sec:res-dr,sec:discussion}).
\end{enumerate}

\paragraph{Relation to our earlier study.} This article reuses the transaction-execution approach of our cost study of classical schemes~\cite{sie2026cost} (ECDSA, P-256, BLS and identity-based signatures), which is available on request to editors and referees. The PQ schemes, all designs and variants, the LMS/HSS verifier and all measurements reported here are new.

\section{Background}
\label{sec:background}

\subsection{Post-quantum signatures}
\emph{ML-DSA}~\cite{ducas2018dilithium,fips204} is a lattice scheme with fast signing and verification and large keys and signatures (1,312 and 2,420 bytes for ML-DSA-44, NIST security category~2; ML-DSA-65 and -87 reach categories 3 and 5). \emph{Falcon}~\cite{falcon2020} has smaller signatures (656 bytes in the compressed encoding for Falcon-512, category~1) but signs with floating-point Gaussian sampling. \emph{SLH-DSA}~\cite{bernstein2019sphincs,fips205} relies only on hash functions (7,856 to 17,088 bytes at category~1). The \emph{stateful} hash-based schemes LMS~\cite{rfc8554} and XMSS~\cite{rfc8391}, approved in SP~800-208~\cite{sp800208}, have small public keys, moderate signatures and hash-only verification; with SHA-256 ($n=32$) they correspond to category~5 and with SHA-256/192 ($n=24$) to category~3. Each one-time key may be used only once, so the signer must keep a counter reliably~\cite{mcgrew2016state}; a tree of height $h$ yields $2^h$ signatures, and HSS stacks trees to extend this. Because the schemes sit at different categories, our comparisons are between the parameter sets a deployment would choose, and \cref{tab:device,tab:stateful} list the category of each.

\subsection{Ethereum gas rules}
A transaction pays 21,000 gas of intrinsic cost, plus calldata, plus execution. A zero calldata byte counts as one token and a non-zero byte as four. Under EIP-7623~\cite{eip7623},
\begin{equation}
G_{\mathrm{tx}}=21{,}000+\max\big(4T+E,\;10T\big),
\label{eq:gas}
\end{equation}
where $T$ is the number of tokens and $E$ the execution gas. When the floor $10T$ binds, a non-zero byte costs 40 gas and additional execution is free as long as $E<6T$. EIP-7825~\cite{eip7825} caps every transaction at $2^{24}$ gas. The Glamsterdam upgrade~\cite{eip7773}, live on the Sepolia testnet since October 2026 and targeted for mainnet in the fourth quarter of 2026, schedules EIP\=7976, which raises the floor to 64 gas per byte for zero and non-zero bytes alike~\cite{eip7976}, together with resource-based intrinsic gas (EIP-2780~\cite{eip2780}), separately metered state-creation gas (EIP-8037~\cite{eip8037}) and higher state-access costs (EIP-8038~\cite{eip8038}). Blob data (EIP-4844~\cite{eip4844}) has a reserve price of 1/16 of the execution base fee per blob gas (EIP-7918~\cite{eip7918}); contracts can read only a blob's commitment, and consensus clients keep blob contents for 4,096 epochs, about 18 days. Under EIP-4444~\cite{eip4444}, execution clients need not serve old block bodies or receipts.

\subsection{Proposed PQ precompiles}
EIP-8051~\cite{eip8051} proposes ML-DSA-44 verification at 4,500 gas with a pre-expanded 20,512-byte public key. EIP-8355~\cite{eip8355} takes standard encodings and charges 6,500, 9,000 and 13,500 gas for ML-DSA-44, -65 and -87 plus 6 gas per 32-byte message word. EIP-8052~\cite{eip8052} splits Falcon-512 verification into hash-to-point (1,000 gas) and core (2,000 gas) precompiles. All three are drafts and none is scheduled for an upgrade; no precompile has been proposed for SLH-DSA, LMS or XMSS. P-256 verification is available since Osaka (EIP-7951~\cite{eip7951}, 6,900 gas).

\section{Related work}
\label{sec:related}

Fern\'andez-Caram\'es and Fraga-Lamas~\cite{fernandez2020towards} survey the exposure of blockchains to quantum attacks. Allende et al.~\cite{allende2023quantum} added a Falcon-512 signature to transactions of an Ethereum-based permissioned network and compared verification by contracts, a modified opcode and precompiles. Kysil et al.~\cite{kysil2025poqeth} implement W-OTS+, XMSS, SPHINCS+ and MAYO verifiers in the EVM with Keccak instead of the standardized hash functions, report 222,114 gas for W-OTS+ ($w=4$), 4.36 million gas for XMSS and 11.6 million gas for SPHINCS+, and evaluate an optimistic ``naysayer'' mode in which signatures are only checked when challenged; they also note that a STARK proof costs about 5 million gas to verify on Ethereum. A SPHINCS variant with limited-signature parameters has been reported at about 150,000 gas~\cite{sphincsminus2026}. PQ transaction signing via account abstraction and precompiles is under discussion~\cite{pqeth2025precompiles}, with signature aggregation identified as the binding constraint~\cite{efpq2026aggregation}; Virani~\cite{virani2026pqsig} derives analytical gas models for verification in smart accounts; and ZKNOX provides Solidity verifiers for Falcon and ML-DSA~\cite{zknox_ethfalcon,zknox_ethdilithium}. These works target transaction authorization or single verifications and do not quantify costs under the calldata floor and transaction cap. Our LMS/HSS verifier differs from the hash-based verifiers above in being compliant with RFC~8554 and SP~800-208 (SHA-256 and SHA-256/192), so that it verifies signatures from standard signers and hardware.

Aguilar-Melchor et al.~\cite{aguilarmelchor2024batch} formalize \emph{batch signatures}, in which one signature authenticates a Merkle tree over many messages, and show that they let slow PQ signers reach high throughput in TLS. Our device-signed roots are an instance of this construction; we study its on-chain consequences. McGrew et al.~\cite{mcgrew2016state} discuss state management for hash-based signatures. On the device side, pqm4~\cite{kannwischer2019pqm4,pqm4bench} benchmarks PQ schemes on Cortex-M4 microcontrollers, Campos et al.~\cite{campos2020lms} compare LMS and XMSS on a Cortex-M4, Pornin~\cite{pornin2019falcon,pornin2026falcon} presents constant-time Falcon implementations for it, and liboqs~\cite{stebila2016oqs} provides portable implementations. Our earlier study~\cite{sie2026cost} found batched ECDSA with Merkle anchoring to be the cheapest classical design; ECDSA and P-256 in the same design serve as baselines here.

\section{System model, threat model and designs}
\label{sec:designs}

\paragraph{Setting.} IoT devices, such as the camera units of an agricultural monitoring system~\cite{sie2022secured}, produce 116-byte records. An untrusted aggregator collects them and submits them to an anchoring contract on Ethereum; submission is permissionless, so a device or any other party can submit a signed root itself, and an aggregator cannot censor it. A \emph{registration authority} (RA) registers each device's public key; the contract stores a 32-byte commitment to it and, optionally, the key itself. The contract has no administrator or upgrade key.

\paragraph{Hashing and encodings.} All on-chain hashing uses Keccak-256. A record leaf is $H(\texttt{0x00}\Vert\mathit{salt}\Vert\mathit{record})$, where an optional per-record salt prevents brute-forcing low-entropy sensor records from published leaves; internal nodes are $H(\texttt{0x01}\Vert\ell\Vert r)$~\cite{rfc6962}, and an odd node is promoted unchanged to the next level. The batch tree over the accepted device roots uses the same rule. Per-record signatures bind $(\mathit{chainid},\allowbreak\mathit{contract},c_{\mathit{dev}},\allowbreak\mathit{idx},s,\allowbreak\mathit{leaf})$ with a per-device sequence number $s$, and device-signed roots bind
\begin{equation}
M=H(\mathit{chainid}\,\Vert\,\mathit{contract}\,\Vert\,c_{\mathit{dev}}\,\Vert\,\mathit{idx}\,\Vert\,e\,\Vert\,R\,\Vert\,k),
\label{eq:msg}
\end{equation}
with key commitment $c_{\mathit{dev}}$, epoch $e$, Merkle root $R$ and record count $k$. The contract accepts $s$ and $e$ only if they exceed the device's last accepted value.

\paragraph{Threat model.} The adversary controls the aggregator and the network, may at some future time run a quantum computer, and then knows every classical key whose public key has been revealed. The contract must accept only records signed by registered devices.
\begin{itemize}[leftmargin=*,nosep]
\item \emph{Registration.} Registrations are authorized by an RA post-quantum signature over $(\mathit{chainid},\allowbreak\mathit{contract},\allowbreak\texttt{"register"},\allowbreak\mathit{idx},\allowbreak H(pk))$ rather than by the RA's ECDSA account. We evaluate ML-DSA-44 and LMS for the RA key; with LMS, the RA is stateful too, but registrations are rare. Revoking or rotating a device key and rotating the RA's own key require messages signed with the current RA key, which is set once at deployment.
\item \emph{Replay and liveness.} Monotonic $s$ and $e$ prevent replay across batches, contracts and chains. Because $e$ only has to increase, a withheld root leaves a visible gap but does not block later roots. For LMS the epoch can be the leaf index $q$; for two-level HSS it is the global index $q_{\mathrm{top}}2^{h_{\mathrm{bottom}}}+q_{\mathrm{bottom}}$, which is monotonic across levels.
\item \emph{State rollback.} A one-time key can be reused after a power loss before the counter is persisted, after restoring a backup, cloning a device or reflashing firmware. Devices should keep the counter in a monotonic hardware counter or secure element, advance it by a safety margin at boot, and never clone key material; if the epoch is the leaf index, the contract also refuses a second root signed with the same index.
\item \emph{Residual dependencies.} The aggregator's transaction signature remains ECDSA; a quantum adversary could forge it to submit or withhold batches, but not to make the contract accept unsigned records. Anchored history ultimately depends on Ethereum's consensus (BLS) signatures, which are outside our scope.
\end{itemize}

\begin{figure}[t]
\centering
\includegraphics[width=\linewidth]{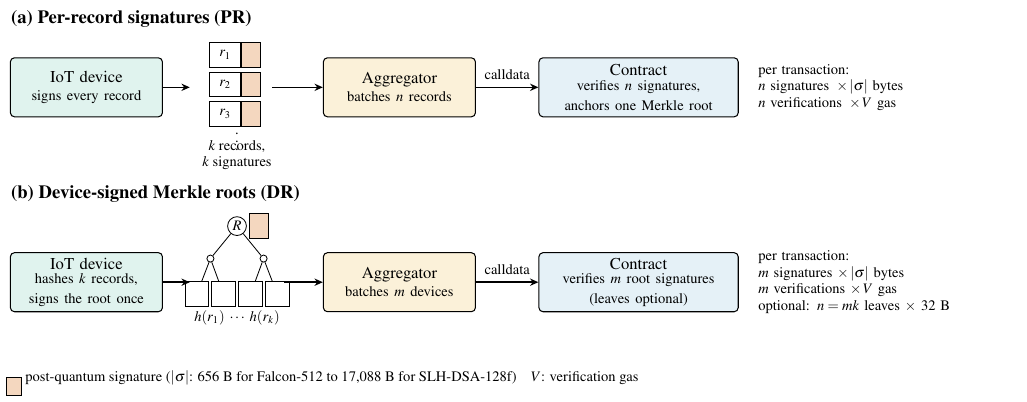}
\caption{(a) Per-record signatures (PR): one signature per record. (b) Device-signed Merkle roots (DR): each device signs one root over its $k$ records, and the contract verifies one signature per device. DR-L posts the leaf hashes in calldata, DR-B in a blob, and DR-0 and DR-0C post only the signed root.}
\label{fig:designs}
\end{figure}

\paragraph{Designs.} With \emph{per-record signatures} (PR), each record is signed and verified individually. With \emph{device-signed Merkle roots} (DR), each device signs $M$ over the root of its $k$ records of an epoch, and the contract verifies one signature per device and stores the root over all accepted device roots of the transaction, so that any record can be proven on-chain with a two-level Merkle path. We evaluate four DR variants (\cref{fig:designs}): \emph{DR-L} posts the $k$ leaves, from which the contract recomputes $R$; \emph{DR-0} posts only $(R,k,e)$, the signature and the key, and logs $(\mathit{idx},e,R,k)$ in an event; \emph{DR-0C} is DR-0 with the key stored once on-chain as contract code (SSTORE2) and read with \texttt{EXTCODECOPY}; and \emph{DR-B} is DR-0 with the leaves in a blob whose versioned hash the contract records. For ECDSA, the ``on-chain key'' is the 20-byte address, since the key itself is recovered from the signature.

\paragraph{Data availability.} The variants differ in what is published and for how long it is guaranteed to remain retrievable (\cref{tab:da}). Only contract storage is permanent. Calldata and event logs are both part of the chain's history, which execution clients may prune under EIP-4444~\cite{eip4444}; beyond that horizon, both depend on archive nodes, indexers or the parties' own copies. The distinction between DR-L and DR-0 is therefore one of \emph{publication}, not retention: DR-L publishes every leaf to all observers at submission time, so anyone can keep a copy, whereas with DR-0 the leaves are never published and only the parties hold them.

\begin{table}[t]
\centering
\caption{Data-availability tiers of the designs.}
\label{tab:da}
\small
\resizebox{\linewidth}{!}{%
\begin{tabular}{@{}llll@{}}
\toprule
Tier & Content in our designs & Guaranteed retention & Who must keep it beyond \\
\midrule
Contract storage & batch roots, epochs, key commitments, stored keys & permanent (state) & nobody \\
Calldata (block bodies) & DR-L leaves; signatures and keys & published to all at submission; prunable under EIP-4444 & archive nodes, indexers, parties \\
Event logs (receipts) & device roots $(\mathit{idx},e,R,k)$ & as calldata & as calldata \\
Blobs & DR-B leaves & about 18 days & parties, blob archivers \\
Off-chain & records, images; DR-0 leaves & none & parties \\
\bottomrule
\end{tabular}}
\end{table}

\paragraph{Cost model.} For a DR transaction with $m$ devices, \cref{eq:gas} gives per-record costs of
\begin{align}
\text{DR-L:}\quad c(k)&\approx A/k+L, \qquad A=21{,}000/m+V+c_b(|\sigma|+|pk|)+g_{\mathrm{dev}},\label{eq:drl}\\
\text{DR-0:}\quad c(k)&\approx A/k,\label{eq:dr0}
\end{align}
where $V$ is the verification gas, $c_b$ the calldata price per byte (16, or 40 when the floor binds), $|pk|=0$ for DR-0C (whose key read is part of $g_{\mathrm{dev}}$), $g_{\mathrm{dev}}$ the per-device overhead (commitment \texttt{SLOAD}, 2,100 gas when cold; epoch update; key hash; event; encoding), and $L$ the cost of posting and hashing one leaf. The relative premium of a PQ scheme over ECDSA in DR-L is then $(A_{\mathrm{PQ}}-A_{\mathrm{EC}})/(A_{\mathrm{EC}}+kL)$, which falls below a target $x$ for
\begin{equation}
k\;\ge\;k^{*}(x)=\frac{A_{\mathrm{PQ}}-(1+x)A_{\mathrm{EC}}}{x\,L},
\label{eq:breakeven}
\end{equation}
whereas in DR-0 the premium is the constant ratio $A_{\mathrm{PQ}}/A_{\mathrm{EC}}$. \Cref{sec:res-dr} fits $A$ and $L$ to the measurements.

\paragraph{Stateful schemes under DR.} DR reduces signing to one signature per device per epoch. With daily roots, an LMS tree of height 10 lasts 1,024 days (2.8 years), height 15 about 90 years, and a two-level HSS of two height-10 trees about 2,900 years. The device persists its counter once per epoch rather than once per record: with daily roots that is 365 writes per year, against about 52,000 for a record every ten minutes, a difference that matters for embedded flash rated for $10^4$--$10^5$ erase cycles per sector. We use LMS rather than XMSS because an LMS chain step costs one hash, whereas an XMSS chain step also derives a key and a bitmask, which roughly triples the hashing in verification~\cite{campos2020lms}.

\section{Methodology}
\label{sec:method}

\paragraph{Device side.} Stateless schemes are timed with liboqs 0.16.0~\cite{stebila2016oqs} C benchmarks (mean of 1\,s runs) on an x86-64 Xeon core, ECDSA with libsecp256k1 and P-256 with OpenSSL; the liboqs Python binding gave results within 10\% of the C figures. The same Python benchmark was run on an ARM64 Neoverse N2 core of a GitHub-hosted runner, through a workflow included in the repository. Stateful schemes are timed with liboqs's C benchmark. Cortex-M4 figures are taken from pqm4~\cite{pqm4bench} (commit 6ff2be9), Campos et al.~\cite{campos2020lms} (LMS, HSS and XMSS on an STM32F4 with SHA-256, reference implementations without tree caching) and Pornin~\cite{pornin2019falcon,pornin2026falcon} (Falcon); times are given at 168\,MHz.

\paragraph{Verifiers.} We reproduce the ZKNOX verifiers~\cite{zknox_ethdilithium,zknox_ethfalcon} with Foundry 1.8.7~\cite{foundry} using each repository's settings and test vectors (ETHDILITHIUM commits f9f4fe1 and f85ca80, ETHFALCON 43ac863). Our LMS/HSS verifiers use the SHA-256 precompile: one specialized for $n=32$ and a generalized one for $n\in\{24,32\}$, both validated against an independent implementation~\cite{hsslms} on valid, tampered and wrong-message inputs. The generalized verifier costs about 21\% more than the specialized one for $n=32$; we therefore compare $n=24$ with $n=32$ within the generalized verifier. LMS verification walks each Winternitz chain from the signed digit to its end, so its cost depends on the message; we regress the gas of 40 LMS-H10-W4 signatures on the number of chain steps and evaluate the fit at the maximum (\cref{app:worst}).

\paragraph{Transactions.} Signed EIP-1559 transactions run on the EthereumJS VM 10.1.3~\cite{ethereumjs} with the Osaka hardfork (EIP-7623, EIP-7825); contracts use Solidity 0.8.28 (optimizer 200 runs, via-IR). Calldata contains genuine keys and signatures. ECDSA is verified for real with \texttt{ecrecover} and LMS/HSS with our verifiers. ML-DSA, Falcon, SLH-DSA, P-256 and hybrid ECDSA+PQ signatures are priced by substitution: the contract calls an empty account with the full input, so that all costs other than verification are measured, including building the input in memory; we subtract the cost of these calls (2,600 gas for the first, cold, access and 100 for each later one) and add, per verification, the verification gas plus a 100-gas call per precompile invoked, since precompiles are always warm under EIP-2929~\cite{eip2929} and charge for their input only through their price. Hybrid signatures carry an ECDSA signature appended to the PQ signature and are charged the PQ price plus \texttt{ecrecover}. Solidity-verifier costs from Foundry already include the cold access, ABI encoding and the callee's memory. Transaction gas is recomputed with \cref{eq:gas}, which matched the VM on all 2,021 stub transactions. Ten LMS signatures measured in isolation predicted a ten-device transaction carrying the same signatures within 0.07\% of its real execution.

\paragraph{Configuration.} All measurements use the revised contract with replay protection and the relaxed epoch rule of \cref{sec:designs}. Every device submits once before measurement, so that counter updates are steady-state writes. Batches contain 10 devices, except DR-L at $k=1{,}000$, which uses 4 devices because 10 exceed the cap. The largest batch per transaction is found by bisection on actual transactions; for PR, records are split evenly over 10 devices.

\section{Results}
\label{sec:results}

\subsection{Device-side costs}
\label{sec:res-device}

\begin{figure}[t]
\centering
\includegraphics[width=\linewidth]{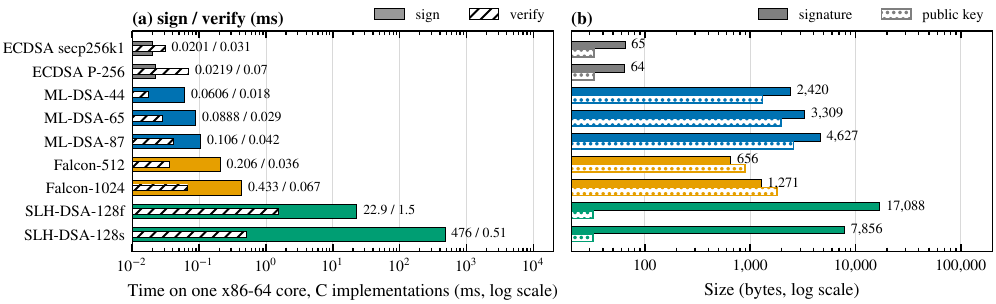}
\caption{Stateless schemes on one x86-64 core (C implementations: liboqs, libsecp256k1, OpenSSL). (a) Mean signing and verification time; labels give sign / verify in ms. (b) Signature and public-key sizes.}
\label{fig:device}
\end{figure}

\begin{table}[t]
\centering
\caption{Stateless schemes on the device side. x86-64: C implementations, mean. ARM64: liboqs Python binding, median. Cortex-M4: published average signing cycles at 168\,MHz, with the implementation used. Cat.: NIST category.}
\label{tab:device}
\small\setlength{\tabcolsep}{3pt}
\resizebox{\linewidth}{!}{%
\begin{tabular}{@{}lcrrrrlrrr@{}}
\toprule
& & \multicolumn{2}{c}{x86-64 (ms)} & \multicolumn{2}{c}{ARM64 (ms)} & \multicolumn{4}{c}{Cortex-M4} \\
\cmidrule(lr){3-4}\cmidrule(lr){5-6}\cmidrule(l){7-10}
Scheme & Cat. & Sign & Verify & Sign & Verify & Implementation & Sign (Mcyc.) & at 168\,MHz & Stack / code (KB) \\
\midrule
ECDSA secp256k1 & -- & 0.020 & 0.031 & 0.028 & 0.042 & -- & -- & -- & -- \\
ECDSA P-256 & -- & 0.022 & 0.070 & -- & -- & -- & -- & -- & -- \\
ML-DSA-44 & 2 & 0.061 & 0.018 & 0.091 & 0.041 & pqm4 m4f & 3.94 & 23\,ms & 43.8 / 19.1 \\
 & & & & & & pqm4 m4fstack & 12.13 & 72\,ms & 5.0 / 24.3 \\
ML-DSA-65 & 3 & 0.089 & 0.029 & 0.130 & 0.067 & pqm4 m4fstack & 24.42 & 145\,ms & 6.5 / 23.6 \\
ML-DSA-87 & 5 & 0.106 & 0.042 & -- & -- & -- & -- & -- & -- \\
Falcon-512 & 1 & 0.207 & 0.036 & 0.174 & 0.028 & Pornin~\cite{pornin2026falcon} & 13.45 & 80\,ms & RAM $\approx$11 \\
Falcon-1024 & 5 & 0.433 & 0.067 & -- & -- & -- & -- & -- & -- \\
SLH-DSA-128f & 1 & 22.9 & 1.53 & 19.6 & 1.20 & pqm4 & 368.6 & 2.2\,s & 2.4 / 4.8 \\
SLH-DSA-128s & 1 & 476 & 0.51 & 407 & 0.41 & pqm4 & 7{,}658 & 45.6\,s & 2.5 / 5.1 \\
\bottomrule
\end{tabular}}
\end{table}

\begin{table}[t]
\centering
\caption{Stateful hash-based schemes. x86-64: liboqs 0.16.0 (C). Cortex-M4: Campos et al.~\cite{campos2020lms}, SHA-256, reference implementations without tree caching (HSS there with two trees of height 5). Signature sizes include the 4-byte HSS header; plain LMS signatures are 4 bytes shorter.}
\label{tab:stateful}
\small\setlength{\tabcolsep}{3.5pt}
\resizebox{\linewidth}{!}{%
\begin{tabular}{@{}lcrrrrrrr@{}}
\toprule
& & & \multicolumn{3}{c}{x86-64} & \multicolumn{2}{c}{Cortex-M4} & \\
\cmidrule(lr){4-6}\cmidrule(lr){7-8}
Scheme & Cat. & Signatures & Key gen. & Sign & Verify & Sign & Verify & Signature (B) \\
\midrule
LMS H10, $w=4$ ($n=32$) & 5 & $2^{10}$ & 0.49\,s & 0.50\,s & 0.24\,ms & 22.6\,s & 16\,ms & 2,512 \\
LMS H10, $w=8$ ($n=32$) & 5 & $2^{10}$ & 3.7\,s & 3.7\,s & 1.70\,ms & 180\,s & 133\,ms & 1,456 \\
LMS H15, $w=4$ ($n=32$) & 5 & $2^{15}$ & 15.1\,s & 15.0\,s & 0.23\,ms & -- & -- & 2,672 \\
HSS, two levels, $w=4$ ($n=32$) & 5 & $2^{20}$ / $2^{10}$ & 0.47\,s & 0.99\,s & 0.47\,ms & 1.6\,s & 18\,ms & 5,076 / 4,756 \\
LMS H10, $w=4$ ($n=24$) & 3 & $2^{10}$ & -- & -- & -- & -- & -- & 1,504 \\
XMSS-SHA2\_10\_256 & 5 & $2^{10}$ & 2.6\,s & 10.5\,ms & 1.32\,ms & 141\,s & 78\,ms & 2,500 \\
\bottomrule
\end{tabular}}
\end{table}

\Cref{fig:device,tab:device,tab:stateful} summarize the device side. On an x86-64 core, ML-DSA signs in 0.06--0.11\,ms, a few times slower than ECDSA, and verifies faster than P-256. The relative cost depends on the architecture: on ARM64, ML-DSA-44 signing is 1.5 times slower than on x86-64, whereas Falcon-512 and SLH-DSA are faster. On a Cortex-M4, ML-DSA-44 signs in 23--72\,ms depending on the stack budget, Falcon-512 in about 80\,ms with 11\,KB of RAM, and SLH-DSA-128f and -128s in 2.2 and 46\,s. Per-record signing at one record every ten minutes would be feasible even for SLH-DSA; the argument for root signing on devices is energy and uplink volume, not throughput.

For the stateful schemes, signing cost depends strongly on the implementation and tree shape. Reference LMS implementations without tree caching recompute the tree for every signature: 0.5\,s for height 10 on the desktop and 22.6\,s on a Cortex-M4, and 3 minutes with $w=8$. A two-level HSS of height 5+5 signs in 1.6\,s on the Cortex-M4, and traversal algorithms with cached nodes, used by liboqs's XMSS (10.5\,ms on the desktop), avoid the recomputation altogether. For daily roots, even 22.6\,s of computation per day is small, but a device should use HSS with small subtrees or a caching implementation.

DR also reduces what the device sends: an ML-DSA-44 device transmits 2,536 bytes per record with per-record signatures, against 140 bytes for $k=100$ and 118 bytes for $k=1{,}000$ with DR, an 18- to 22-fold reduction; the device keeps the Merkle frontier, $(\lceil\log_2 k\rceil+1)\cdot32$ bytes (352 bytes for $k=1{,}000$), or $32k$ bytes if it buffers leaf hashes.

\subsection{Verification costs}
\label{sec:res-solidity}

\begin{table}[t]
\centering
\caption{Verification gas per signature. Solidity ML-DSA and Falcon: published values next to our reproduction (``bench'': repository benchmark vector; ``KAT'': known-answer test). LMS/HSS: our verifiers. Precompiles: proposed or deployed prices.}
\label{tab:verify}
\small
\resizebox{\linewidth}{!}{%
\begin{tabular}{@{}llrr@{}}
\toprule
Scheme & Implementation & Published & Reproduced / measured \\
\midrule
ML-DSA-44 & Solidity, NIST (ZKNOX main / optimized branch) & 8.1M / 1,196,707 & 8,128,360 / 1,192,929 (KAT) \\
          & Solidity, EVM-friendly (ZKNOX main / optimized branch) & 4.9M / 847,709 & 4,833,650 / 845,438 (KAT) \\
          & Precompile EIP-8355 (6,500 + 6 per message word) / EIP-8051 & 6,506 / 4,500 & -- \\
Falcon-512 & Solidity, NIST-compliant / EVM-friendly (ZKNOX) & -- & 4,818,843 / 1,547,336 (bench) \\
           & Precompile EIP-8052 & 3,000 & -- \\
LMS-H10-W4, $n=32$ & Solidity, specialized (this work) & -- & mean 268,677 (246,817--294,342; 40 signatures); worst case 429,900 \\
LMS-H15-W4, $n=32$ & Solidity, specialized (this work) & -- & median 280,790 (257,155--294,955; 12 signatures) \\
HSS (H10-W4)$^2$, $n=32$ & Solidity, specialized (this work) & -- & median 581,403 (557,297--605,010; 12 signatures) \\
LMS-H10-W4, $n=24$ / $n=32$ & Solidity, generalized (this work) & -- & median 270,743 / 327,684 (6 signatures each) \\
P-256 / ECDSA & Precompile EIP-7951 / \texttt{ecrecover} & 6,900 / 3,000 & -- \\
\bottomrule
\end{tabular}}
\end{table}

\Cref{tab:verify} lists the verification costs. Our reproductions match the published ML-DSA figures within 1.5\% (0.3\% for the optimized branch). NIST-compliant ML-DSA-44 and Falcon-512 cost 1.2 and 4.8 million gas; EVM-friendly variants, which replace SHAKE by Keccak and are not interoperable with standard signers, cost 0.85 and 1.55 million. Our standard-compliant LMS verifier costs about 0.27 million gas on average, four times less than the optimized NIST ML-DSA verifier and sixteen times less than the Keccak-based XMSS verifier of~\cite{kysil2025poqeth}. Its cost is linear in the number of Winternitz chain steps: 111,315 gas plus 322 gas per step ($R^2=0.996$), with 420--570 steps observed and 990 in the worst case, which costs 429,900 gas, 60\% above the mean; aggregators must set gas limits for this case. A larger tree adds little (height 15: +4\%), two-level HSS doubles the cost, and SHA-256/192 ($n=24$, category~3) reduces it by 17\% and the signature by 40\% compared with $n=32$ in the same verifier. A Keccak instantiation would save roughly 140 gas per chain step at the price of interoperability.

\subsection{Per-record signatures}
\label{sec:res-pr}

\begin{table}[t]
\centering
\caption{Per-record signatures (PR) with replay protection: one record per transaction, gas per record in batches of 50 (10 devices, 5 records each), and the largest batch under the cap (records split evenly over 10 devices, e.g.\ $520=10\times52$; $\approx$: estimated from the per-record cost). SLH-DSA rows are lower bounds (data cost only).}
\label{tab:pr}
\small
\begin{tabular}{@{}llrrr@{}}
\toprule
Scheme & Verification & Individual & Per record, $n=50$ & Max.\ records/tx \\
\midrule
ECDSA secp256k1 & \texttt{ecrecover} & 64,675 & \textbf{12,971} & 1,430 \\
ML-DSA-44 & Solidity, NIST (optimized) & 1,313,896 & $>$cap & $\approx$13 \\
          & Solidity, EVM-friendly (optimized) & 966,405 & $>$cap & $\approx$18 \\
          & Precompile EIP-8355 & 177,660 & 112,490 & 160 \\
ML-DSA-65 & Precompile EIP-8355 & 238,620 & 153,188 & -- \\
ML-DSA-87 & Precompile EIP-8355 & 317,090 & 210,833 & -- \\
Falcon-512 & Solidity, EVM-friendly & 1,633,265 & $>$cap & $\approx$10 \\
           & Precompile EIP-8052 & 91,780 & 39,513 & 520 \\
LMS-H10-W4 & Solidity (this work) & 360,215 & 326,555 & 50 \\
SLH-DSA-128s & lower bound (data only) & $\ge$344,060 & $\ge$319,585 & -- \\
SLH-DSA-128f & lower bound (data only) & $\ge$712,250 & invalid ($>$cap) & -- \\
\bottomrule
\end{tabular}
\end{table}

\begin{figure}[t]
\centering
\includegraphics[width=\linewidth]{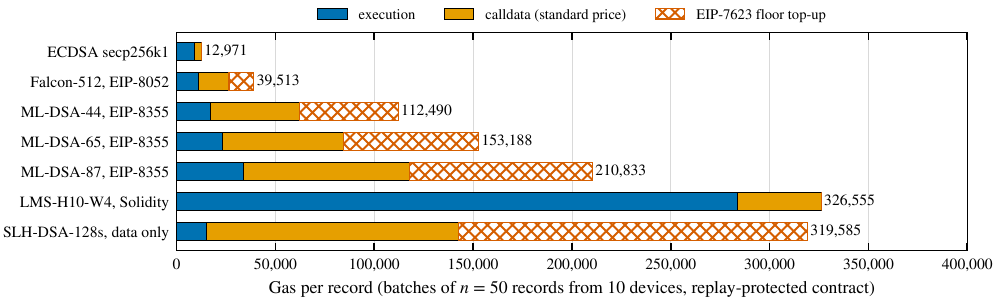}
\caption{Composition of the per-record cost of PR batches ($n=50$, 10 devices, replay-protected contract). Where the EIP-7623 floor binds, the hatched part is its surcharge over the standard calldata price, and execution below $6T$, including verification, adds nothing.}
\label{fig:breakdown}
\end{figure}

\Cref{tab:pr} and \cref{fig:breakdown} report the PR design. With the proposed precompiles, an ML-DSA-44 record costs 112,490 gas (8.7 times ECDSA) and a Falcon-512 record 39,513 gas (3.0 times): the floor binds, every byte of signature and key costs 40 gas, and the verification adds nothing. With today's Solidity verifiers a transaction holds 10 to 50 PQ-signed records. Whether a per-record cost of $10^{-4}$\,ETH is acceptable depends on the application (\cref{sec:res-sens}); at scale, the decisive limit is capacity, about 530 ML-DSA-44 records per 60-million-gas block.

\begin{finding}
Under EIP-7623, per-record PQ authentication is priced by signature and key size, not by verification; cheaper precompiles alone cannot make it competitive with ECDSA.
\end{finding}

\subsection{Device-signed Merkle roots}
\label{sec:res-dr}

\begin{table}[t]
\centering
\caption{DR-L (leaves posted, keys in calldata): gas per record and factor relative to ECDSA in the same design. 10 devices per transaction; $^\dagger$4 devices.}
\label{tab:drl}
\small\setlength{\tabcolsep}{4pt}
\resizebox{\linewidth}{!}{%
\begin{tabular}{@{}llrrrr@{}}
\toprule
Scheme & Verification & $k=10$ & $k=100$ & $k=1{,}000^\dagger$ & vs.\ ECDSA ($k$=100 / 1,000) \\
\midrule
ECDSA secp256k1 & \texttt{ecrecover} & 3,821 & 1,558 & 1,473 & 1 / 1 \\
ECDSA P-256 & EIP-7951 & 4,273 & 1,604 & 1,478 & 1.03 / 1.00 \\
ML-DSA-44 & Precompile EIP-8355 & 16,834 & 2,831 & 1,551 & 1.82 / 1.05 \\
          & Solidity, NIST (optimized) & 128,866 & 14,097 & 2,738 & 9.0 / 1.86 \\
ML-DSA-65 & Precompile EIP-8355 & 22,942 & 3,443 & 1,586 & 2.2 / 1.08 \\
Falcon-512 & Precompile EIP-8052 & 8,227 & 1,972 & 1,505 & 1.27 / 1.02 \\
           & Solidity, EVM-friendly & 160,724 & 17,267 & 3,050 & 11.1 / 2.07 \\
Hybrid ECDSA + ML-DSA-44 & EIP-8355 + \texttt{ecrecover} & 17,170 & 2,867 & 1,556 & 1.84 / 1.06 \\
Hybrid ECDSA + Falcon-512 & EIP-8052 + \texttt{ecrecover} & 8,547 & 2,003 & 1,510 & 1.29 / 1.03 \\
LMS-H10-W4, $n=32$ & Solidity (this work) & 34,607 & 4,675 & 1,791 & 3.0 / 1.22 \\
LMS-H10-W2, $n=32$ & Solidity (this work) & 38,747 & 5,047 & 1,829 & 3.2 / 1.24 \\
LMS-H10-W8, $n=32$ & Solidity (this work) & 152,155 & 16,481 & 2,928 & 10.6 / 1.99 \\
LMS-H10-W4, $n=24$ & Solidity, generalized (this work) & 31,444 & 4,345 & 1,747 & 2.8 / 1.19 \\
SLH-DSA-128s & lower bound (data only) & 33,447 & 4,494 & 1,615 & 2.9 / 1.10 \\
\bottomrule
\end{tabular}}
\end{table}

\begin{table}[t]
\centering
\caption{Root-only designs: gas per record with keys in calldata (DR-0) or stored on-chain (DR-0C), factor relative to the cheaper ECDSA root-only variant ($k=100$), and the largest number of devices per transaction. LMS-H15 and HSS were measured in a separate run with the same contract.}
\label{tab:dr0}
\small\setlength{\tabcolsep}{3.5pt}
\resizebox{\linewidth}{!}{%
\begin{tabular}{@{}llrrrrrr@{}}
\toprule
& & \multicolumn{2}{c}{DR-0 (key in calldata)} & \multicolumn{2}{c}{DR-0C (key on-chain)} & vs.\ ECDSA & Max.\ devices/tx \\
Scheme & Verification & $k$=100 & $k$=1,000 & $k$=100 & $k$=1,000 & ($k$=100) & (key placement) \\
\midrule
ECDSA secp256k1 & \texttt{ecrecover} & 245 & 25 & 284 & 28 & 1 & 689 (calldata) \\
ECDSA P-256 & EIP-7951 & 290 & 29 & 322 & 32 & 1.18 & 656 (calldata) \\
ML-DSA-44 & Precompile EIP-8355 & 1,555 & 156 & 1,031 & 103 & 4.2 & 166 (on-chain) \\
ML-DSA-44 & Precompile EIP-8051 (expanded key) & -- & -- & 1,142 & 114 & 4.7 & 58 (on-chain) \\
ML-DSA-44 & Solidity, NIST (optimized) & 12,746 & 1,275 & 12,572 & 1,257 & 51 & 13 (on-chain) \\
ML-DSA-65 & Precompile EIP-8355 & 2,167 & 217 & 1,386 & 139 & 5.7 & -- \\
Falcon-512 & Precompile EIP-8052 & 695 & 69 & 387 & 39 & 1.6 & 404 (on-chain) \\
Falcon-512 & Solidity, EVM-friendly & 15,934 & 1,593 & 15,828 & 1,583 & 65 & 10 (on-chain) \\
Hybrid ECDSA + ML-DSA-44 & EIP-8355 + \texttt{ecrecover} & 1,590 & 159 & 1,057 & 106 & 4.3 & 161 (on-chain) \\
Hybrid ECDSA + Falcon-512 & EIP-8052 + \texttt{ecrecover} & 726 & 73 & 428 & 43 & 1.7 & -- \\
LMS-H10-W4, $n=32$ & Solidity (this work) & 3,350 & 335 & 3,455 & 345 & 13.7 & $\approx$50 (calldata) \\
\quad worst case & & 4,963 & 496 & -- & -- & 20.3 & $\approx$33 (calldata) \\
LMS-H10-W4, $n=24$ & Solidity, generalized (this work) & 2,991 & 299 & 3,026 & 303 & 12.2 & 56 (calldata) \\
LMS-H15-W4, $n=32$ & Solidity (this work) & 3,456 & 346 & -- & -- & 14.1 & -- \\
HSS (H10-W4)$^2$, $n=32$ & Solidity (this work) & 6,884 & 688 & -- & -- & 28.1 & -- \\
SLH-DSA-128s & lower bound (data only) & 3,217 & 322 & 3,204 & 320 & 13.1 & -- \\
\bottomrule
\end{tabular}}
\end{table}

\begin{table}[t]
\centering
\caption{Cost model for DR-L, \cref{eq:drl}, fitted on $k=10$ and $k=100$ (10 devices): per-device term $A$, per-leaf term $L$, error of the prediction at $k=1{,}000$, and the smallest $k$ at which the premium over ECDSA falls below 25\% and 10\% (\cref{eq:breakeven}).}
\label{tab:model}
\small
\begin{tabular}{@{}lrrrrr@{}}
\toprule
Scheme / verification & $A$ (gas) & $L$ (gas) & Error at $k$=1,000 & $k^*(25\%)$ & $k^*(10\%)$ \\
\midrule
ECDSA, \texttt{ecrecover} & 25,144 & 1,307 & $-9.4\%$ & -- & -- \\
P-256, EIP-7951 & 29,656 & 1,307 & $-9.3\%$ & 0 & 15 \\
ML-DSA-44, EIP-8355 & 155,589 & 1,275 & $-7.6\%$ & 389 & 1,003 \\
ML-DSA-44, Solidity NIST & 1,275,211 & 1,345 & $-4.2\%$ & 3,699 & 9,276 \\
Falcon-512, EIP-8052 & 69,500 & 1,277 & $-10.3\%$ & 119 & 328 \\
Hybrid ECDSA + ML-DSA-44 & 158,922 & 1,278 & $-7.5\%$ & 399 & 1,027 \\
LMS-H10-W4, $n=32$ & 332,578 & 1,349 & $-5.9\%$ & 893 & 2,260 \\
LMS-H10-W4, $n=24$ & 301,100 & 1,334 & $-6.2\%$ & 809 & 2,050 \\
SLH-DSA-128s (data only) & 321,700 & 1,277 & $-0.8\%$ & 909 & 2,303 \\
\bottomrule
\end{tabular}
\end{table}

\begin{figure}[t]
\centering
\includegraphics[width=\linewidth]{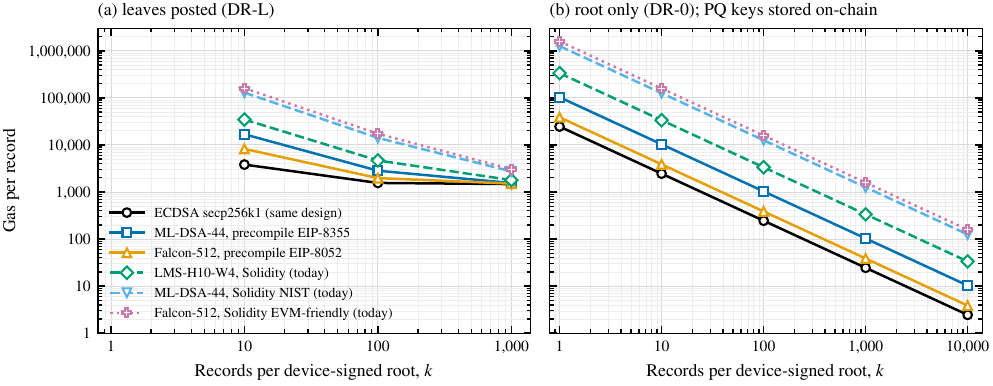}
\caption{Device-signed roots: gas per record versus records per signature $k$. (a) DR-L, leaves posted (10 devices; 4 for $k=1{,}000$). (b) Root only, each scheme with its cheaper key placement (PQ schemes with precompiles: DR-0C; ECDSA and LMS: DR-0). ECDSA is shown in the same design in both panels.}
\label{fig:dr}
\end{figure}

\Cref{tab:drl,tab:dr0,tab:model} and \cref{fig:dr} report the device-root designs, each against ECDSA in the same design. With leaves posted (DR-L), the schemes with precompiles and LMS approach the cost of one leaf at $k=1{,}000$: ML-DSA-44 with a precompile costs 5\% more than ECDSA (82\% more at $k=100$), Falcon-512 2\%, and our LMS verifier 22\% (19\% with $n=24$). With only roots posted (DR-0, DR-0C), costs fall as $1/k$ and the relative premium is larger, from 1.6 times ECDSA (Falcon-512 with precompile) to 65 times (Falcon-512 in Solidity), while absolute costs stay small: at $k=100$, 1,031 gas per record for ML-DSA-44 with precompile and on-chain key, 3,350 gas for LMS (4,963 in the worst case) and 2,991 gas for LMS with $n=24$. P-256, the curve of most IoT secure elements, costs 3\% (DR-L) to 18\% (DR-0) more than secp256k1. A hybrid ECDSA and PQ signature, as recommended for the transition period, adds at most 2\% in DR-L and 2.5\% (ML-DSA-44) to 11\% (Falcon-512) in DR-0C over the PQ signature alone.

The cost model of \cref{eq:drl} predicts the 4-device, $k=1{,}000$ measurements within 0.4--10.5\%, underestimating them slightly because memory costs grow superlinearly in very large transactions. It gives a direct rule for choosing $k$: for the premium over ECDSA to fall below 25\% with leaves posted, Falcon-512 with a precompile needs about 120 records per root, ML-DSA-44 about 390, LMS about 800--900, and NIST ML-DSA in Solidity about 3,700.

\begin{finding}
Device-signed roots make post-quantum authentication of IoT records deployable on Ethereum today. A standard LMS verifier in Solidity costs 3,350 gas per record (4,963 in the worst case) at $k=100$ when only roots are posted, and 1,791 gas at $k=1{,}000$ when every record hash is published; for LMS, the premium over ECDSA in the same design ranges from 19\% to about 20 times, depending on the variant, $k$ and hash length.
\end{finding}

\subsection{Key registration and storage}
\label{sec:res-keys}

\begin{table}[t]
\centering
\caption{Registration gas (one-time per device).}
\label{tab:reg}
\small
\resizebox{\linewidth}{!}{%
\begin{tabular}{@{}lrrr@{}}
\toprule
Scheme & Commitment (key in calldata) & Key stored on-chain & PQ-authorized by RA signature \\
\midrule
ML-DSA-44 & 74,570 & 383,973 & 171,650 (precompile) / 1,305,054 (Solidity) \\
ML-DSA-65 & 100,200 & 522,755 & -- \\
Falcon-512 & 59,499 & 294,125 & -- \\
ML-DSA-44, EIP-8051 key & 840,830 & 4,550,102 & -- \\
LMS-H10-W4 & 45,589 & 111,635 & 371,525 (Solidity, real) \\
\bottomrule
\end{tabular}}
\end{table}

\Cref{tab:reg} lists registration costs. Storing an ML-DSA-44 key on-chain costs 309,403 gas more than registering its commitment and saves about 52,000 gas per root submission, paying off after six submissions; it also makes EIP-8051's 20,512-byte expanded keys usable (114,142 gas per device root). Under Glamsterdam, EIP-8037 meters state creation separately and EIP-8038 raises state-access costs~\cite{eip8037,eip8038}, which will make on-chain key storage, registration and the per-batch root write more expensive; the break-even shifts accordingly.

\subsection{Sensitivity, rollups and fees}
\label{sec:res-sens}

\begin{table}[t]
\centering
\caption{Gas per record under the current floor, the Glamsterdam floor (EIP-7976, scheduled) and a hypothetical further increase, recomputed with \cref{eq:gas} from the measured execution and byte counts; and calldata bytes per record, the quantity a rollup posts to Ethereum. PR at $n=50$, DR at $k=100$.}
\label{tab:sens}
\small\setlength{\tabcolsep}{4pt}
\begin{tabular}{@{}llrrrr@{}}
\toprule
Design & Scheme / verification & 10/40 (now) & 64/64 (Glamsterdam) & 96/96 & Bytes/record \\
\midrule
PR & ECDSA & 12,971 & 24,264 & 36,186 & 373 \\
PR & ML-DSA-44, precompile & 112,490 & 190,152 & 285,018 & 2,965 \\
PR & Falcon-512, precompile & 39,513 & 72,597 & 108,685 & 1,128 \\
PR & LMS-W4, Solidity & 326,555 & 326,555 & 326,555 & 2,811 \\
DR-L & ECDSA & 1,558 & 2,381 & 3,560 & 36.9 \\
DR-L & ML-DSA-44, precompile & 2,831 & 4,695 & 7,032 & 73.0 \\
DR-L & LMS-W4, Solidity & 4,675 & 4,675 & 5,926 & 61.5 \\
DR-0 & ECDSA & 245 & 333 & 488 & 4.9 \\
DR-0 & LMS-W4 ($n=32$ / $n=24$), Solidity & 3,350 / 2,991 & 3,350 / 2,991 & 3,350 / 2,991 & 29.5 / 19.3 \\
DR-0C & ML-DSA-44, precompile & 1,031 & 1,807 & 2,700 & 27.9 \\
DR-0C & Falcon-512, precompile & 387 & 681 & 1,011 & 10.3 \\
\bottomrule
\end{tabular}
\end{table}

\paragraph{Calldata repricing.} Under the Glamsterdam floor, floor-bound designs become 1.4 to 1.9 times more expensive, whereas LMS is unaffected (\cref{tab:sens}); LMS's premium over ECDSA falls from 3.0 to 2.0 times with leaves posted and from 13.7 to 10.1 times with roots only. A further increase to 96/96 would raise floor-bound designs by 2.0--2.8 times and, for the first time, LMS with posted leaves (+27\%). Glamsterdam's other repricings change the intrinsic cost (EIP-2780), state creation (EIP-8037, outside the transaction cap) and state access (EIP-8038), which affect registration, key storage and the per-batch storage write but not the leaf and signature costs that dominate our results.

\paragraph{Blob cost.} A blob holds 4,096 field elements of 31 usable bytes, i.e.\ 3,968 leaves, and transactions pay for whole blobs. At the EIP-7918 reserve price, a blob costs the equivalent of 8,192 gas, so DR-B would add about 2 gas per leaf for full blobs, 8 for a 1,000-leaf batch and 82 for a 100-leaf batch, multiplied by the ratio of the market to the reserve blob price. These are estimates; we did not execute blob transactions.

\paragraph{Rollups.} On a rollup the dominant cost is the data posted to Ethereum, which compression cannot reduce for high-entropy signatures. For ML-DSA-44, per-record signatures post 2,965 bytes per record against 28--73 bytes with device-signed roots, a 40- to 100-fold reduction (10 bytes for Falcon-512 with DR-0C, 19--30 for LMS). DR-0 with LMS and DR-0C with ML-DSA-44 post almost the same data (29.5 and 27.9 bytes); on a rollup where execution is cheap, the choice of scheme then nearly disappears and Solidity verifiers become competitive.

\paragraph{Fees and capacity.} A record costs its gas times the base fee plus the priority fee. At 0.1, 1 and 10\,gwei base fee plus 0.1\,gwei priority, a per-record ML-DSA-44 signature with precompile costs $2.3\times10^{-5}$, $1.2\times10^{-4}$ and $1.1\times10^{-3}$\,ETH, LMS with root signing ($k=100$) $6.7\times10^{-7}$, $3.7\times10^{-6}$ and $3.4\times10^{-5}$\,ETH, and ML-DSA-44 with precompile and on-chain keys $2.1\times10^{-7}$, $1.1\times10^{-6}$ and $1.0\times10^{-5}$\,ETH. With the 60-million block gas limit at the time of writing (November 2026)~\cite{eip7935}, a block could anchor about 4,600 records with per-record ECDSA, 530 with per-record ML-DSA-44 and about 58,000 with device-signed ML-DSA-44 roots ($k=100$), ignoring other traffic. Glamsterdam is expected to enable higher gas limits, which scale these capacities but not the per-transaction results.

\section{Discussion}
\label{sec:discussion}

\paragraph{When is on-chain verification worth its premium?} Two alternatives avoid verifying device signatures in the contract. \emph{Proof-based aggregation}: an aggregator proves that it verified $m$ device signatures, and the contract verifies one proof. Under a PQ requirement the proof cannot be wrapped in a pairing-based SNARK, the usual way to shrink on-chain cost, so a STARK must be verified directly, at about 5 million gas~\cite{kysil2025poqeth}; with a proof of the order of 100 kilobytes in calldata, a proof costs some 6--9 million gas. Against DR-0C with ML-DSA-44 (about 103,000 gas per device root) this breaks even at roughly 60--90 devices per proof, and against LMS (about 335,000 gas) at roughly 18--27 devices; beyond that, aggregation is cheaper on-chain, at the price of a prover, its hardware and its latency. \emph{Optimistic anchoring}: the contract records each device's root and a hash of its signature without verifying it, and verifies only when challenged~\cite{kysil2025poqeth}; this costs roughly the ECDSA root-only figure per device, about 25,000 gas, independent of the PQ scheme, but requires watchers, a challenge window during which records are not final, and availability of the signatures during that window. Contract-level verification is worth its premium when other contracts must consume records immediately and trustlessly, for example to release a loan instalment when a production milestone is anchored, without relying on provers, watchers or challenge periods; at small fleet sizes it is also the cheapest of the three.

\paragraph{Choosing a design.} If every record hash should be published to all observers at submission, DR-L with $k$ chosen from \cref{tab:model} brings all schemes close to the cost of a leaf. If the parties can keep the leaves themselves, DR-0 is an order of magnitude cheaper. A camera recording every ten minutes produces 144 records per day, so daily roots correspond to $k\approx150$. Root signing adds latency: with daily roots, a record waits up to a day before it is authenticated on-chain, and records buffered on a device that fails before signing are lost unless the device forwards them immediately and signs their hashes later. Applications that need records on-chain within minutes can combine ECDSA per record with a periodic PQ root over the already anchored record hashes, at the DR-0 cost for the PQ layer.

\paragraph{Which scheme today.} Without precompiles, LMS is the cheapest standard-compliant option, about four times cheaper to verify than the optimized NIST ML-DSA verifier, robust to calldata repricing, and with $n=24$ (category~3) 11\% cheaper still in DR-0. Its price is statefulness: the counter must be persisted and protected against rollback, and the tree must be sized for the device's lifetime, which HSS extends at twice the verification cost. ML-DSA is stateless and simpler on devices; Falcon has the smallest signatures and, with a precompile, the lowest on-chain cost, but its signing is harder to implement securely on microcontrollers. SLH-DSA is a conservative fallback; a SPHINCS variant with limited-signature parameters has been verified on-chain for about 150,000 gas~\cite{sphincsminus2026}, but it is not the standardized parameter set, so we report only data costs. A hybrid ECDSA and PQ signature costs little more than the PQ signature alone and is a sensible default during the transition.

\paragraph{Limitations.} No measurement was taken on IoT hardware: device figures come from x86-64 and ARM64 server cores and from published Cortex-M4 results, and energy was not measured. We did not execute blob transactions or measure on a rollup, and the aggregation and optimistic alternatives are analysed, not implemented. Precompile results depend on draft prices, although \cref{fig:breakdown} shows that per-record costs are insensitive to them. The Solidity verifiers are experimental and unaudited, as are our LMS/HSS verifiers. The results reflect the Osaka rules and the repricing scheduled for Glamsterdam; the released harness can recompute them for later changes.

\section{Conclusion}
\label{sec:conclusion}

Signing every IoT record with a post-quantum signature is uneconomical at scale on Ethereum and remains so with the proposed precompiles, because the EIP-7623 floor prices every signature byte, and the scheduled Glamsterdam repricing makes it worse. Signing Merkle roots on the device changes this: with roots over a hundred records, PQ authentication costs about a thousand gas per record with precompiles and stored keys, and a few thousand gas with a standard LMS verifier that runs on Ethereum today and is unaffected by the repricing. These costs exceed those of ECDSA in the same design by a factor that the cost model makes explicit, and they come with batching latency and data-availability properties that must match the application. Devices deployed now can therefore be given a post-quantum authentication path without waiting for protocol changes.

\section*{Data and code availability}
The device-side benchmarks, the anchoring contracts, the LMS/HSS verifiers, the transaction harnesses, the scripts reproducing the Solidity verifier measurements, and all raw results are available at \url{https://github.com/Siemingfong/pq-iot-auth} (release v2.3), archived at Zenodo under DOI \href{https://doi.org/10.5281/zenodo.23970176}{10.5281/zenodo.23970176}; release v2.2, DOI \href{https://doi.org/10.5281/zenodo.23969084}{10.5281/zenodo.23969084}, contains the code of an earlier version of this manuscript.

\section*{Author contributions}
The author is solely responsible for the conceptualization, methodology, software, investigation, formal analysis, visualization and writing of this article.

\section*{Acknowledgements}
The author thanks the ZKNOX team for releasing their Solidity verifiers and the pqm4 contributors for their public benchmarks.

\section*{Funding}
This research did not receive any specific grant from funding agencies in the public, commercial, or not-for-profit sectors.

\section*{Declaration of competing interest}
The author is a co-founder and the chief executive officer of Conzen.AI, a company developing privacy-preserving computation for financial compliance and supply chain traceability. The company did not fund this study and had no role in its design, analysis or reporting.

\section*{Declaration of generative AI and AI-assisted technologies in the research and writing process}
The author used Claude (Anthropic) to assist with writing the benchmark code and Solidity contracts, running the experiments, analysing data, preparing figures, and drafting and editing the text. All code, measurements and text were reviewed by the author, who takes full responsibility for the content of the publication.

\appendix
\section{Worst case of LMS verification}
\label{app:worst}

For LMOTS with $w=4$ and $n=32$, the message digest $Q$ has 64 base-16 digits $a_i$ and the checksum $S=\sum_{i}(15-a_i)\le960$ is shifted left by 4 bits and contributes three further digits $c_1,c_2,c_3$, the hexadecimal digits of $S$. The verifier walks $15-a_i$ steps on each message chain and $15-c_j$ on each checksum chain, i.e.\ $S+45-\mathrm{ds}(S)$ steps in total, where $\mathrm{ds}(S)$ is the hexadecimal digit sum of $S$. Over $0\le S\le960$, this is maximal at $S=960=\texttt{0x3C0}$, with $\mathrm{ds}=15$, giving 990 steps; the 40 measured signatures required 420--570 steps. The regression of verification gas on steps, $111{,}315+322\cdot\mathrm{steps}$ with $R^2=0.996$, gives 429,900 gas at 990 steps.

\bibliographystyle{unsrtnat}
\bibliography{references}
\end{document}